\documentclass[conference,a4paper]{IEEEtran}
\IEEEoverridecommandlockouts
\usepackage{cite}
\usepackage{amsmath,amssymb,amsfonts}
\usepackage{algorithmic}
\usepackage{graphicx}
\usepackage{textcomp}
\def\BibTeX{{\rm B\kern-.05em{\sc i\kern-.025em b}\kern-.08em
    T\kern-.1667em\lower.7ex\hbox{E}\kern-.125emX}}
\usepackage{graphics} 
\usepackage{graphicx}   
\graphicspath{{pic/}}   
\usepackage{subfigure}  
\usepackage[table]{xcolor}
\usepackage{multirow}
\usepackage{float}
\usepackage[section]{placeins}
\usepackage{booktabs}
\usepackage{threeparttable}
\usepackage{url}

\begin{document}

\title{Intra Encoding Complexity Control with a Time-Cost Model for Versatile Video Coding}
\author{
Yan Huang$^{1}$, Jizheng Xu$^{3}$, Li Zhang$^{3}$, Yan Zhao$^{1}$, Li Song$^{1, 2}$\\
$^{1}$Institute of Image Communication and Network Engineering, Shanghai Jiao Tong University\\
$^{2}$MoE Key Lab of Artificial Intelligence, AI Institute, Shanghai Jiao Tong University\\
$^{3}$ByteDance Inc.\\
E-mail: 2250344200@sjtu.edu.cn, song\_li@sjtu.edu.cn (corresponding author)\\
} 
\maketitle

\begin{abstract}
For the latest video coding standard Versatile Video Coding (VVC), the encoding complexity is much higher than previous video coding standards to achieve a better coding efficiency, especially for intra coding. The complexity becomes a major barrier of its deployment and use. Even with many fast encoding algorithms, it is still practically important to control the encoding complexity to a given level. Inspired by rate control algorithms, we propose a scheme to precisely control the intra encoding complexity of VVC. In the proposed scheme, a Time-PlanarCost (viz. Time-Cost, or T-C) model is utilized for CTU encoding time estimation. By combining a set of predefined parameters and the T-C model, CTU-level complexity can be roughly controlled. Then to achieve a precise picture-level complexity control, a framework is constructed including uneven complexity pre-allocation, preset selection and feedback. Experimental results show that, for the challenging intra coding scenario, the complexity error quickly converges to under 3.21\%, while keeping a reasonable time saving and rate-distortion (RD) performance. This proves the efficiency of the proposed methods.
\end{abstract}

\begin{IEEEkeywords}
VVC, complexity, control, fast, partition
\end{IEEEkeywords}

\section{Introduction}
Versatile Video Coding (VVC) \cite{vvcdraft}, developed by a joint collaborative team of ITU-T and ISO/IEC experts known as the Joint Video Experts Team (JVET), 
has a compression capability that is substantially higher than its predecessor, i.e., High Efficiency Video Coding (HEVC) \cite{sullivan2012overview}.
A more sophisticated algorithm design, as described in \cite{vtm}, brought about around 40\% of bitrate saving, but also led to significant encoding complexity increase. 
According to \cite{siqueira2020rate}, although Single Instruction Multiple Data is integrated and enabled in VVC reference encoder implementation, VVC encoding time is still on average 10.2 times higher than HEVC's under Random Access (RA) setting. As for All Intra (AI) setting, the complexity is even increased by a factor of 31\cite{pakdaman2020complexity}. Encoding complexity is an obvious barrier for wide deployment and use of VVC. For this reason, reduction of encoding complexity has already become a hot research topic even before the completion of VVC standard. Currently most of the research focuses on complexity reduction of some modules, for example, skipping full search of quadtree with nested multi-type tree (QTMT) partition process\cite{amestoy2019tunable}, fast Intra mode decision \cite{yang2019low}, and fast multiple transform selection \cite{fu2019two}. Actually a series of complexity reduction algorithms are also discussed during the JVET development, some of which have been adopted in the VVC test model (VTM), as presented in \cite{wieckowski2019fast}.

However, complexity reduction algorithms may not solve the problem in practice. There are two reasons. First, complexity reduction algorithms ordinarily only have several discrete configurations. So it is not flexible enough to meet the demand of a resource constraint application whose target encoding time is between two configurations. Second, the performance of complexity reduction is not stable. For the same algorithm, a huge difference of time saving and RD loss could appear on different sequences. 
On the other hand, complexity control is a must to fill the gap between demands and fast encoding algorithms.
There have been some attempts to explore complexity control methods on HEVC \cite{huang2020online, zhang2017ctu, huang2021modeling}. But for those methods, only acceleration by a given ratio instead of a specific target time realization is supported. To realize precise encoding time control, the original total encoding time should be collected at first. On the other hand, for VVC, whose partitioning process is much more complicated and complexity control is more difficult to achieve accordingly. As far as we know, currently there is no related research that has been presented.

\begin{figure}[tb]
\centering
 \subfigure[Time-PlanarCost]{
 \includegraphics[width=0.45\linewidth]{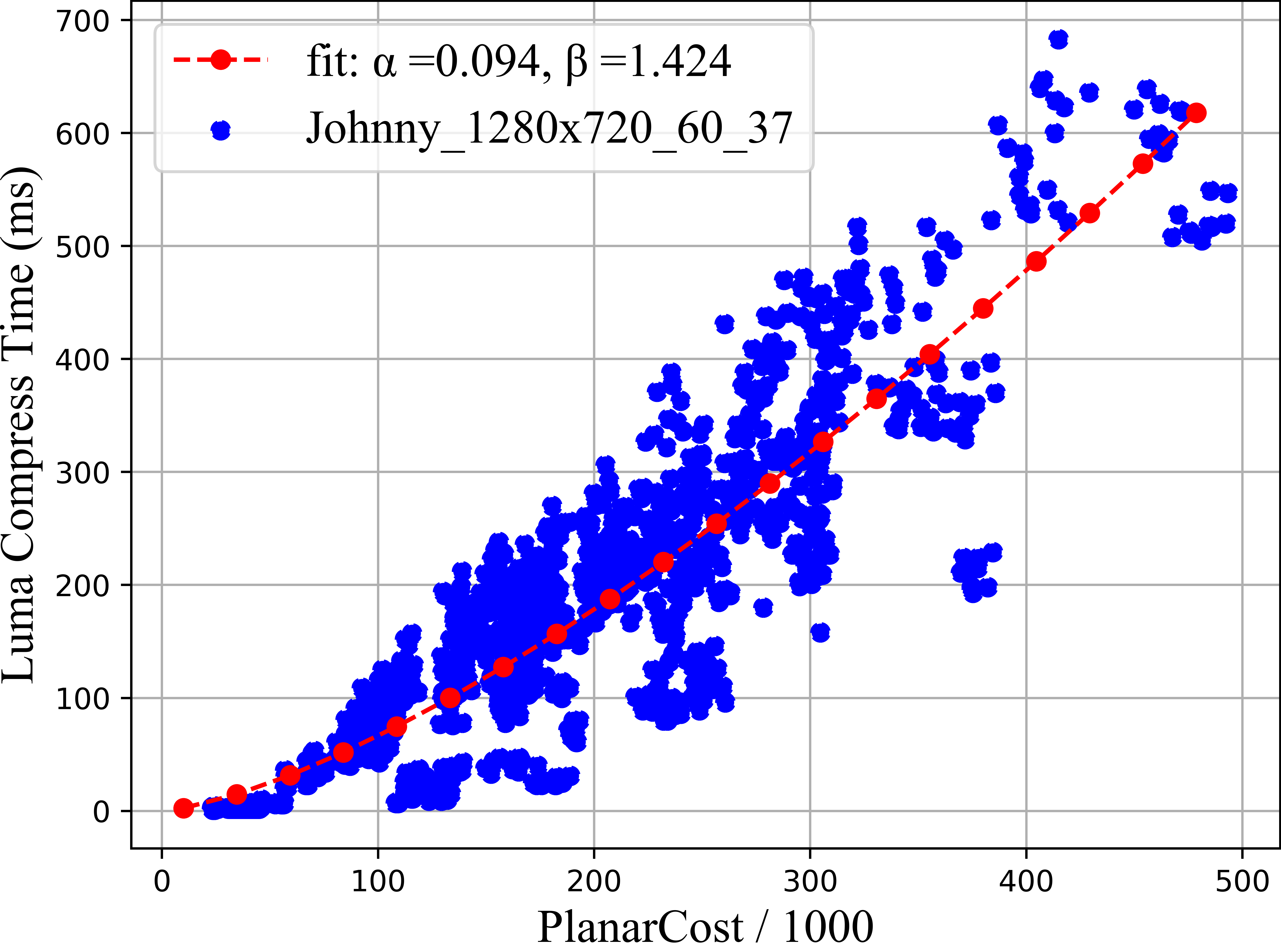}
 \label{fig:model-a}
 }
 \subfigure[Time Prediction]{
 \includegraphics[width=0.45\linewidth]{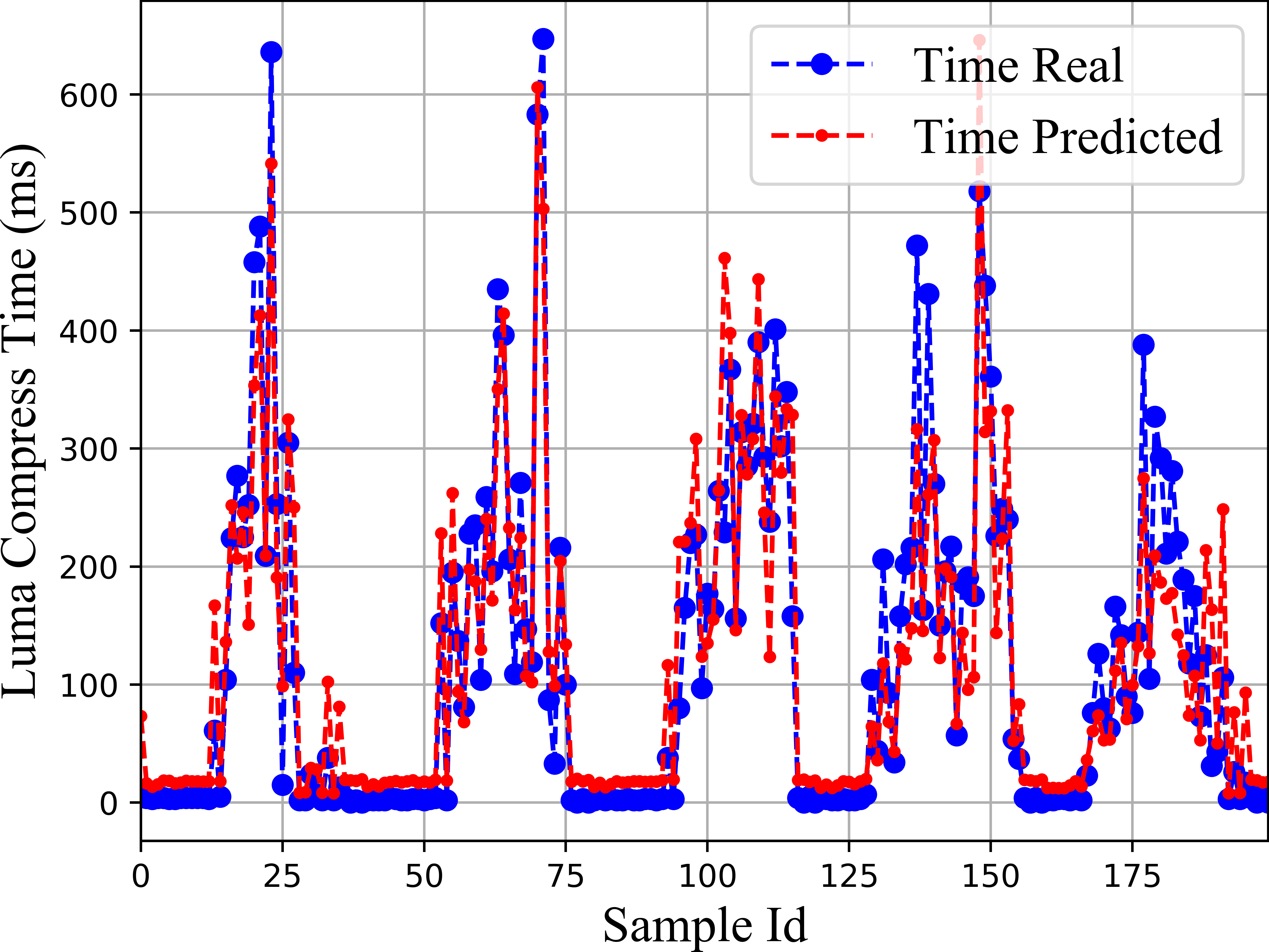}
 \label{fig:model-b}
 }
\caption{The performance of the T-C model on "Johnny" under QP 37}
  \label{fig:model}
\vspace{-0.4cm}
\end{figure}

Therefore, the specific objective of this study is to investigate a one-pass complexity (more specifically, encoding time) control method for VVC Intra coding. 
To guarantee the controllability of encoding time, and also to make the framework simple and clear, we choose to control the most important component of CTU time, i.e., Luma compress time, which composes more than $80\%$ of total encoding time.
A CTU-level Time-PlanarCost (viz. T-C) model is used for CTU Luma compress time estimation, followed by an efficient preset constrained QTMT depth selection to reduce complexity. By combining these two methods presented in Section II, a target encoding time of each CTU can be roughly realized. To guarantee the accuracy of picture-level encoding time. A framework that contains techniques including uneven complexity pre-allocation, encoding preset selection and feedback is further proposed, which is described in Section III. Experimental results are illustrated in Section IV. Then, Section V concludes the paper.


\section{Model and Presets}
In order to achieve the target encoding time of a single CTU, two critical problems should be considered in advance.
First, an accurate method to estimate CTU encoding time under default configuration (i.e., original CTU time) should be designed. Only when the original CTU time is precisely estimated can we convert the CTU time target to acceleration ratio.
Second, under the assumption that we have precisely acquired acceleration ratio, how to approximate it with as low rate-distortion (RD) loss as possible.
In this section, we propose a PlanarCost-based model to estimate original CTU time and design presets, i.e., predefined parameters, to approximate the acceleration ratio based on Pareto rules.
Considering the maximal CU size of AI setting is 64, the approximation is conducted at 64 block level, which provides finer grain for encoding time adjustment. For convenience, we still use CTU to denote those 64 CU blocks.

\subsection{Time-Cost Model}
To estimate encoding time priors to the real encoding process, Luma compress time of each CTU is collected. Different features are tried and for the corresponding block, PlanarCost, i.e., the RDO cost for luma Planar mode is found to have a strong correlation with its Luma compress time. We take 30 frames of data from "Johnny" under QP 37 as a representative to show the relationship between these two elements in Fig. \ref{fig:model-a}. 
A power relationship is observed, and therefore a Time-PlanarCost (viz. T-C) model is constructed as
\begin{equation}
\label{eq:TDModel}
T_p=\alpha \times PlanarCost^{\beta}
\end{equation}
where $\alpha$ and $\beta$ are sequences and QP-related parameters. Here PlanarCost value is scaled down by a factor of 1000 to avoid a small value of $\alpha$. Then $\alpha$ and $\beta$ in (\ref{eq:TDModel}) are approximated offline. During the approximation process, PlanarCost and Luma compress time of CTUs from the first ten frames are collected and used. Then the approximated parameters, i.e., $\alpha$ and $\beta$ can be used to predict the Luma compress time. Fig. \ref{fig:model-b} demonstrates the predicted and the corresponding real Luma compress time of the first 200 CTUs on Sequence "Johnny". It can be observed the trend of real Luma compress time (blue curve) can be well traced by predicted one (red curve), which proves the accuracy of the model.

With the help of the accurate model, once PlanarCost is available after Intra mode search finishes, the original Luma compress time of each CTU can be immediately estimated. 
Then, by further combining the CTU budget, the acceleration ratio can be derived to control the subsequent partition process. 

\subsection{Data Driven Presets}
To realize the acceleration ratio derived in the last part, we need to design a set of parameters (or presets) with different acceleration abilities. A carefully designed fast algorithm, for example, CNN and SVM-based algorithm \cite{galpin2019cnn, wu2021svm} could have better performance, but we choose to constrain QTMT maximum depth directly for two reasons. On the one hand, to highlight the framework's strengths and avoid confusion from the advantages of fast algorithms, we should simplify the design of our presets. On the other hand, according to our preliminary experiment, directly constraining the QTMT depth is direct and effective.

We traverse the combinations of QT with maximum depths of $\{0,1,2,3,4\}$, BT of $\{0,1,2,3,4,5\}$ and MT of $\{0,1,2,3\}$. All sequences from Class C-E are tested to investigate the acceleration properties as regards to time saving (TS) and RD loss (evaluated by Bj{\o}ntegaard delta bit-rate (BDBR) \cite{bjontegaard2001calculation} , also shorted as BR). 
The frontier presets, which means the best cost performance are extracted to chosen configurations in the TABLE \ref{tab:presets} as our data driven presets $0\sim5$.

\begin{table}[tb]
  \centering
  \caption{The Chosen Frontier Presets}
    \begin{tabular}{cccccc}
    \toprule
    \multirow{2}[4]{*}{\textbf{Preset}} & \multicolumn{3}{c}{\textbf{Maximum Depth}} & \multirow{2}[4]{*}{\textbf{BDBR (\%)}} & \multirow{2}[4]{*}{\textbf{TS}} \\
\cmidrule{2-4}          & \textbf{QT} & \textbf{BT} & \textbf{MT} &       &  \\
    \midrule
    \textbf{0} & /     & /     & /     & 0.00  & 0.00\% \\
    \textbf{1} & 4     & 4     & 3     & 0.03  & 1.20\% \\
    \textbf{2} & 4     & 3     & 3     & 0.33  & 9.20\% \\
    \textbf{3} & 4     & 3     & 2     & 1.52  & 46.30\% \\
    \textbf{4} & 4     & 2     & 1     & 3.70  & 63.40\% \\
    \textbf{5} & 4     & 0     & 0     & 9.02  & 72.90\% \\
    \bottomrule
    \end{tabular}%
  \label{tab:presets}%
\vspace{-0.4cm}  
\end{table}%

\section{Complexity Control Framework}
\begin{figure}[tb]
\centering
 \includegraphics[width=1\linewidth]{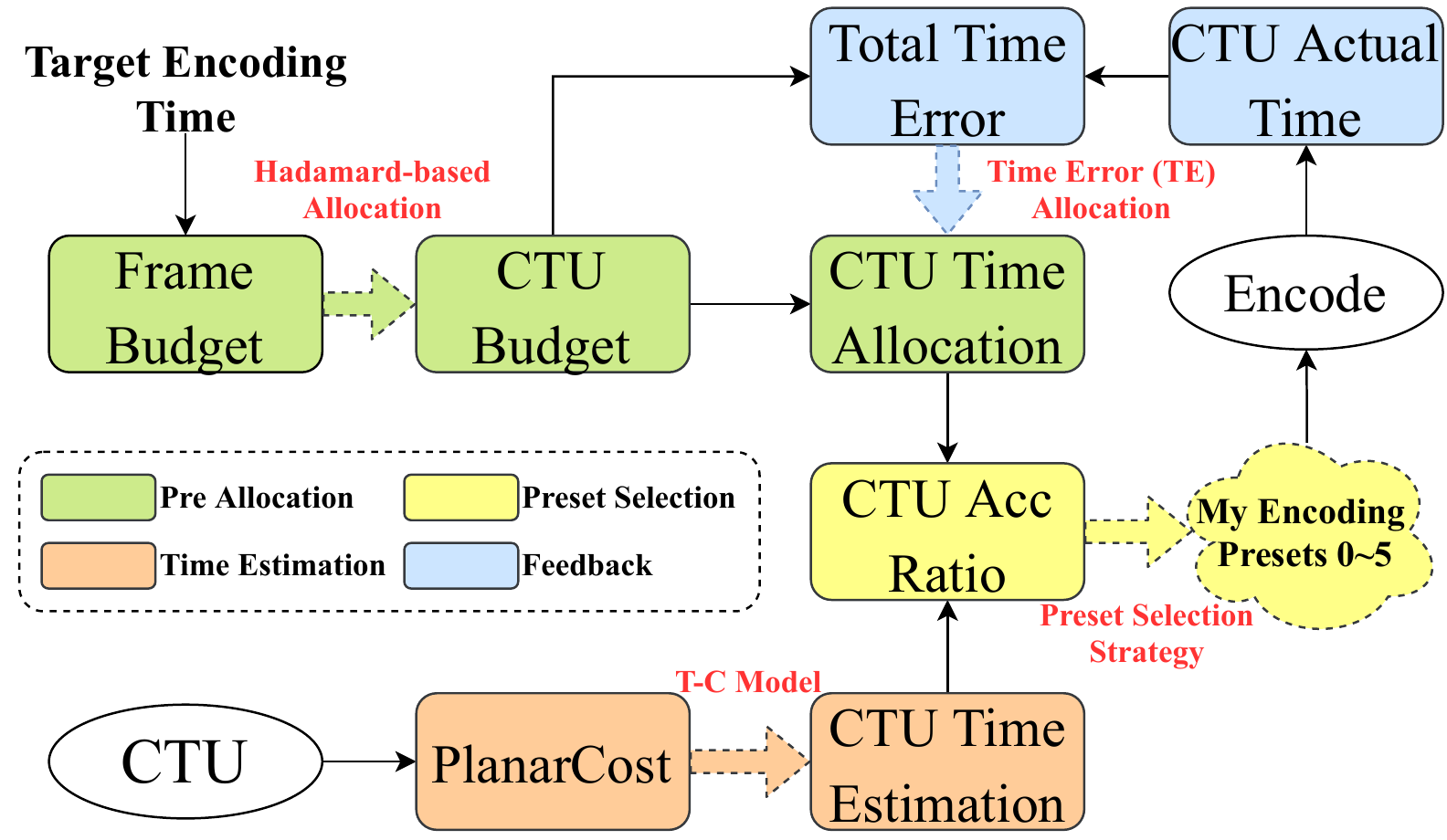}
\caption{The overall complexity control framework}
  \label{fig:framework} 
\vspace{-0.4cm}
\end{figure}

The last section describes a model and some presets, which are key factors for coarse-grain CTU-level budget realization. In this section, given a specific picture-level coding budget, we will introduce the proposed framework to realize picture-level precise complexity control. The overall framework is shown in Fig. \ref{fig:framework}, which is composed of four major components, i.e., complexity pre-allocation (green blocks), complexity estimation (orange blocks), encoding preset selection (yellow blocks) and feedback (blue blocks). Among them, complexity estimation has been introduced in Section II, and the rest components will be illustrated respectively.
\subsection{Complexity Pre-allocation}
Although a precise relationship between PlanarCost and encoding time exists, precise PlanarCost value is not available until the execution of the Intra search for each CTU. For the sake of determining the weight of all CTUs in the whole frame before encoding, we choose to use sum of absolute $8\times8$ Hadamard cost, i.e., SA8D cost, because of the positive correlation with encoding time. Let $H$ be the hadamard matrix and $block$ be the original pixel, the weight is defined as
\begin{equation}
\label{eq:Had}
\omega _{CTU}=\sum_{i\,\,=\,\,1}^{Num\left( 8*8 blocks \right)}{\sum_8{\sum_8{\left| H\times block_i\times H \right|}}}
\end{equation}
Then, the picture-level budget will be unevenly allocated to each CTU according to the corresponding SA8D cost. 
\begin{equation}
\label{eq:TPreA}
T_b=\frac{\omega _{CTU}}{\sum_{Pic}{\omega _{CTU}}}\times T_{PicBudget}
\end{equation}
where $\omega _{CTU}$ means the weight of each CTU and $T_b$ denotes the time budget for it.
The allocation process regards complexity as a factor impacting original CTU encoding time. Through this allocation, CTUs with higher complexity will be allocated more time budget, which will avoid unreasonable acceleration ratio for those complex blocks. 
According to our preliminary results, introducing this uneven complexity pre-allocation will provide around 1\% less BDBR under the same frame budget. Notably, the SA8D costs are also calculated for the rectangle CTUs at the picture boundary, and CTU time budgets are allocated accordingly.

\subsection{Encoding Preset Selection}
The encoding preset selection process is inserted right after Intra prediction of 64 block, when PlanarCost can be accessed and deeper partitions has not been tried yet. The acquired PlanarCost is then plugged into (\ref{eq:TDModel}) to estimate original CTU Luma compress time. To maintain the accurate characteristics of the model on different compiling or running environment, we design a calibration factor $r_{cpu}$ to represent relative computing capability as below
\begin{equation}
\label{eq:rCPU}
r_{cpu}=\frac{\sum_{CTUs\,\,NoAcc}{T_r}}{\sum_{CTUs\,\,NoAcc}{T_p}}
\end{equation}
where $T_r$ means real encoding time collected online, while $T_p$ is the predicted value from our model (\ref{eq:TDModel}). During encoding, CTUs not accelerated ("NoAcc" in equation) will be used to calculate the calibration factor, which will be immediately utilized to update $\alpha$ as
\begin{equation}
\label{eq:Tcpu}
\tilde{\alpha}=r_{cpu}\times \alpha
\end{equation}
This calibration mechanism will make the model more general and applicable, by constantly calibrating model accuracy while encoding. Then the updated $\tilde{\alpha}$ will be used to predict original Luma compress time of the following CTUs as
\begin{equation}
\label{eq:Tp}
\tilde{T}_p=\tilde{\alpha}\times PlanarCost^{\beta}
\end{equation}

On the other hand, CTU budget obtained from (\ref{eq:TPreA}) will be updated with $T_{fb}$, i.e., the time feedback to current CTU from previously accumulated time error.
\begin{equation}
\label{eq:TAlloc}
T_a=T_b+T_{fb}
\end{equation}
Here we name the updated CTU time budget as allocated CTU time $T_a$. Combining allocated CTU time and predicted original CTU time, the target time ratio of the current CTU, i.e., $r_{CTU}$ will be derived as
\begin{equation}
\label{eq:rCTU}
r_{CTU}=\frac{T_a}{\tilde{T}_p}
\end{equation}
Then, the preset in TABLE \ref{tab:presets} closet to $r_{CTU}$ will be adopted to approximate the target acceleration ratio.

\subsection{Complexity Error Feedback}
At the end of the encoding process of each CTU, the actual consumed time of the CTU $T_r$ can be obtained, which can be used to calculate time error $T_e$ as
\begin{equation}
\label{eq:TErrorCTU}
T_e=T_r-T_a
\end{equation}

Before the encoding process of each CTU, the previous accumulated time error will be used to calculate the time error feedback as
\begin{equation}
\label{eq:Tfb}
T_{fb}=\frac{\sum_{CTUs\,\,Coded}{T_e}}{N_{Window}}
\end{equation}
where $N_{Window}$ denotes the number of CTU used for error assignment. Smaller $N_{Window}$ could make time fast convergence, but too strict a time limit would result in a higher RD loss. In this paper, anchor $N_{Window}$ is set as 20, and the number of left CTU $CTUleft$ is also considered for the time error suppression during the end of a frame. 
\begin{equation}
\label{eq:Window}
N_{Window}=\max \left( 1, \min \left( CTUleft, 20 \right) \right) 
\end{equation}
After determining $N_{Window}$, time feedback $T_{fb}$ will be obtained, which is used to update the time budget of the CTU to be encoded as (\ref{eq:TAlloc}).
Through the joint function of the proposed four components, Luma compress time of the whole frame will be allocated, monitored, and adjusted through feedback. By adopting the proposed framework, theoretically, Luma compress time, i.e., the major component of the overall frame encoding time, will be precisely controlled.

\section{Experimental Results}
In this section, we will evaluate the effectiveness of the proposed complexity control framework from two aspects, i.e., per-sequence complexity control accuracy and overall performance under target time. 
The framework is implemented in VTM10.0. Four QP values {22, 27, 32, 37}  were chosen for compressing sequences from Class A1, A2, B, C, E of the JVET standard test set \cite{2018JVET}. Class D is excluded because of the excessive few CTUs in a frame.
The calculation of time saving (TS) over the original VTM10.0 is listed as (\ref{e:ts}).
\begin{equation}
\label{e:ts}
TS=\frac{T_{VTM10.0}-T_{\mathrm{Proposed}}}{T_{VTM10.0}}*100\%
\end{equation}
And time error (TE) is used to calculate the picture-level time error ratio from the time budget of this frame as
\begin{equation}
\label{e:te}
TE=\frac{\left| \sum_{CTUs\,\,in\,\,Pic}{T_r}-T_{PicBudget} \right|}{T_{PicBudget}}*100\%
\end{equation}

\subsection{Per-sequence Complexity Control Accuracy}
In this part, we evaluate the major function, i.e., the complexity control accuracy of our framework, by a challenging task to control the encoding time within a frame. 
Before testing, one frame of each test sequence is encoded by default VTM10.0 encoder, where Luma compress time is collected. 
Then a ratio from 30\% to 90\% will be multiplied to derive target Luma compress time.
Notably, only a specific target time will be given to the redesigned VTM encoder. Neither original time nor acceleration ratio will be used. 
Then through combining the pre-allocated budget and feedback, the encoder will automatically analyze and choose QTMT depths of each CTU to approximate the picture-level complexity budget.

\begin{figure}[tb]
\centering
 \subfigure[Campfire (3840x2016)]{
 \includegraphics[width=0.4\linewidth]{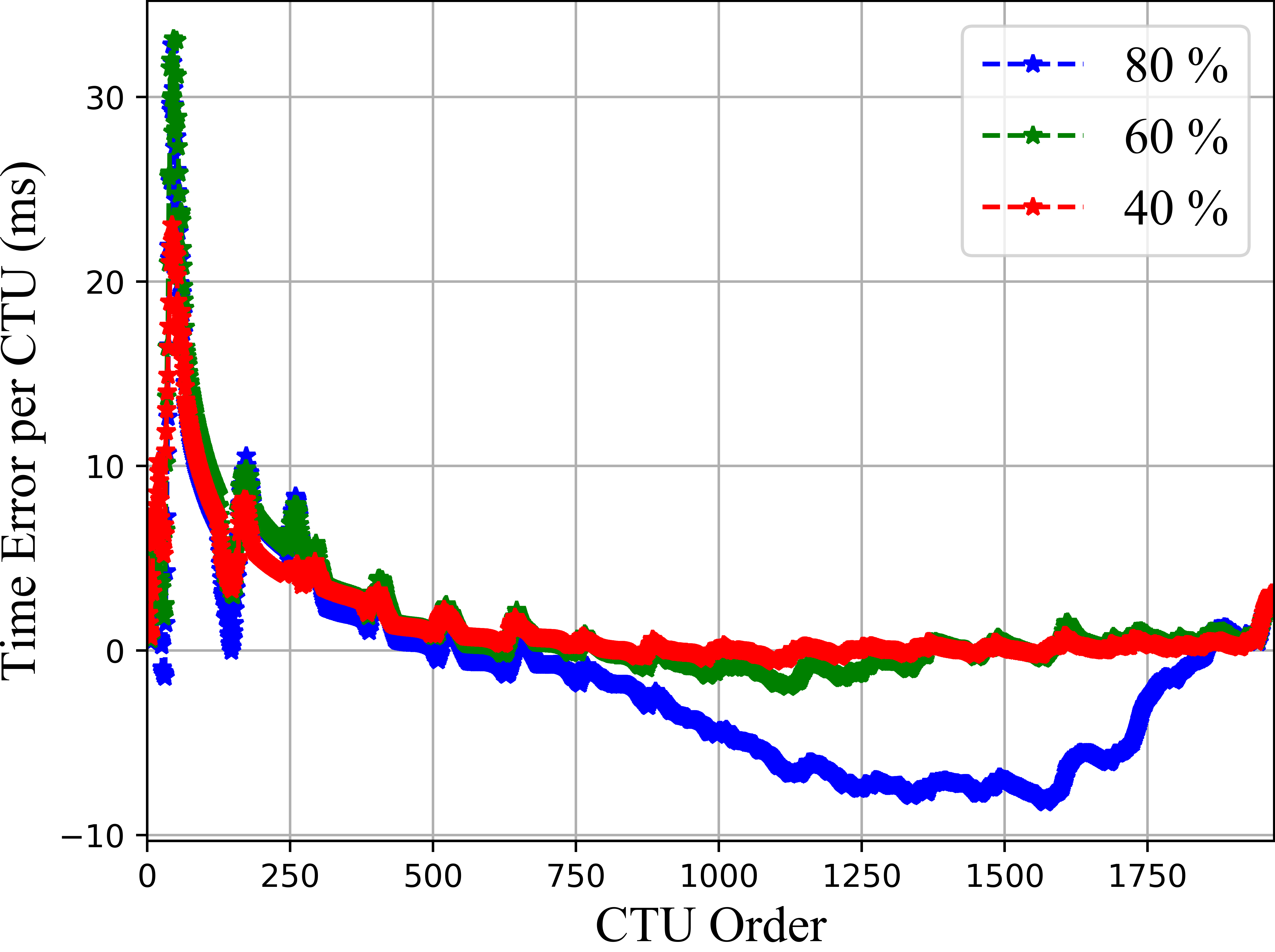}}
 \subfigure[Cactus (1920x1080)]{
 \includegraphics[width=0.4\linewidth]{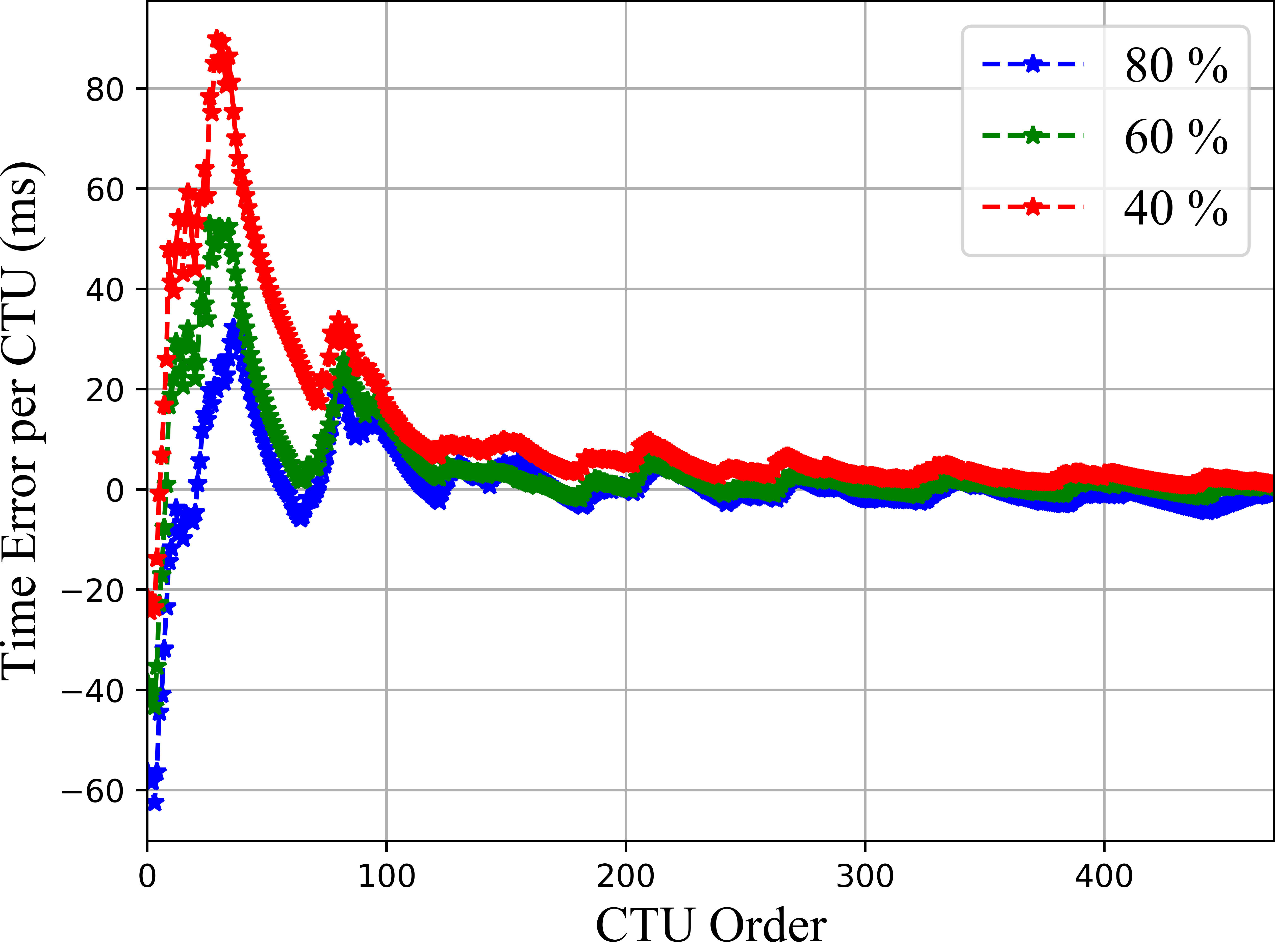}}
 \subfigure[Johnny (1280x720)]{
 \includegraphics[width=0.4\linewidth]{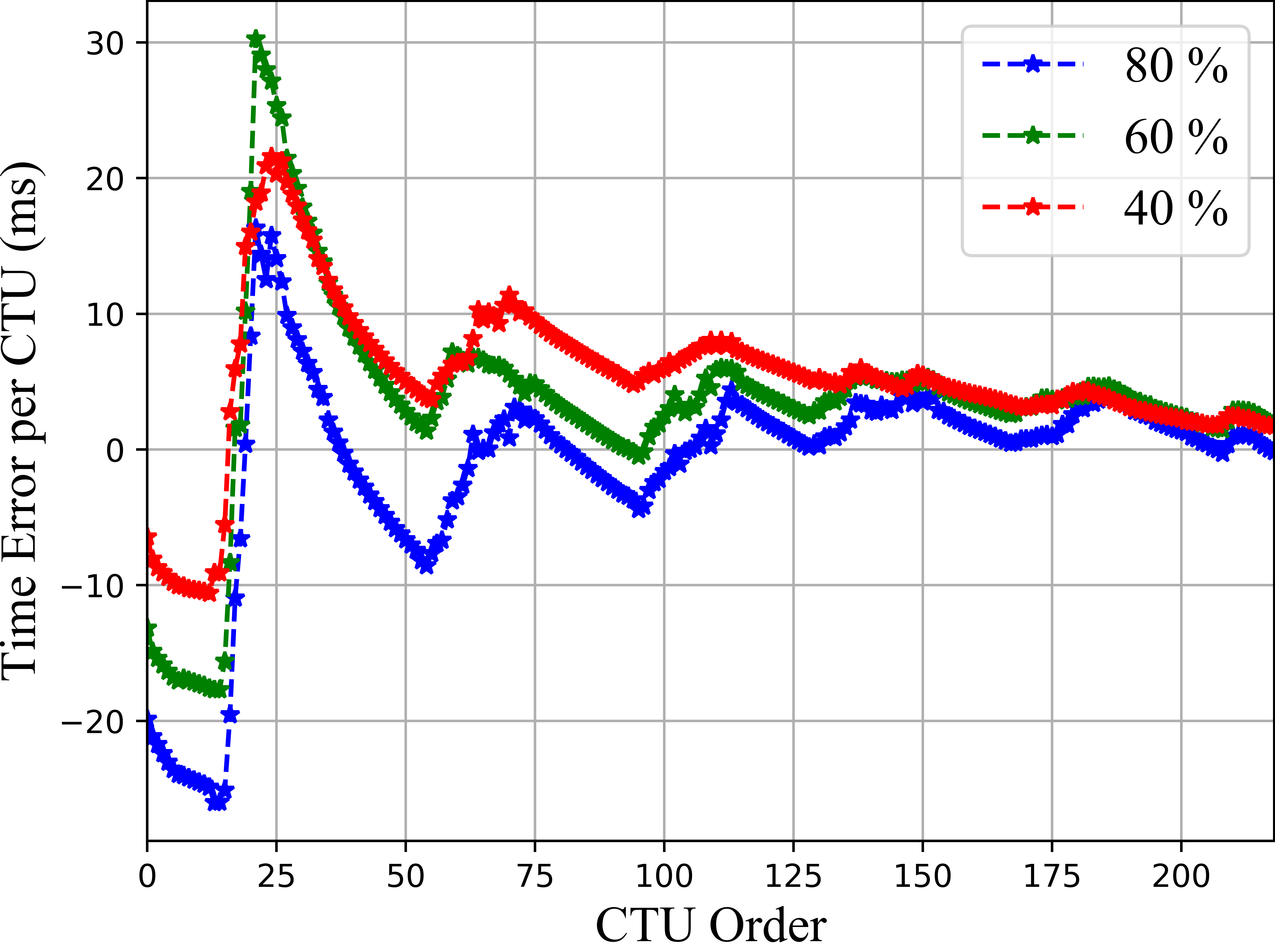}}
 \subfigure[PartyScene (832x480)]{
 \includegraphics[width=0.4\linewidth]{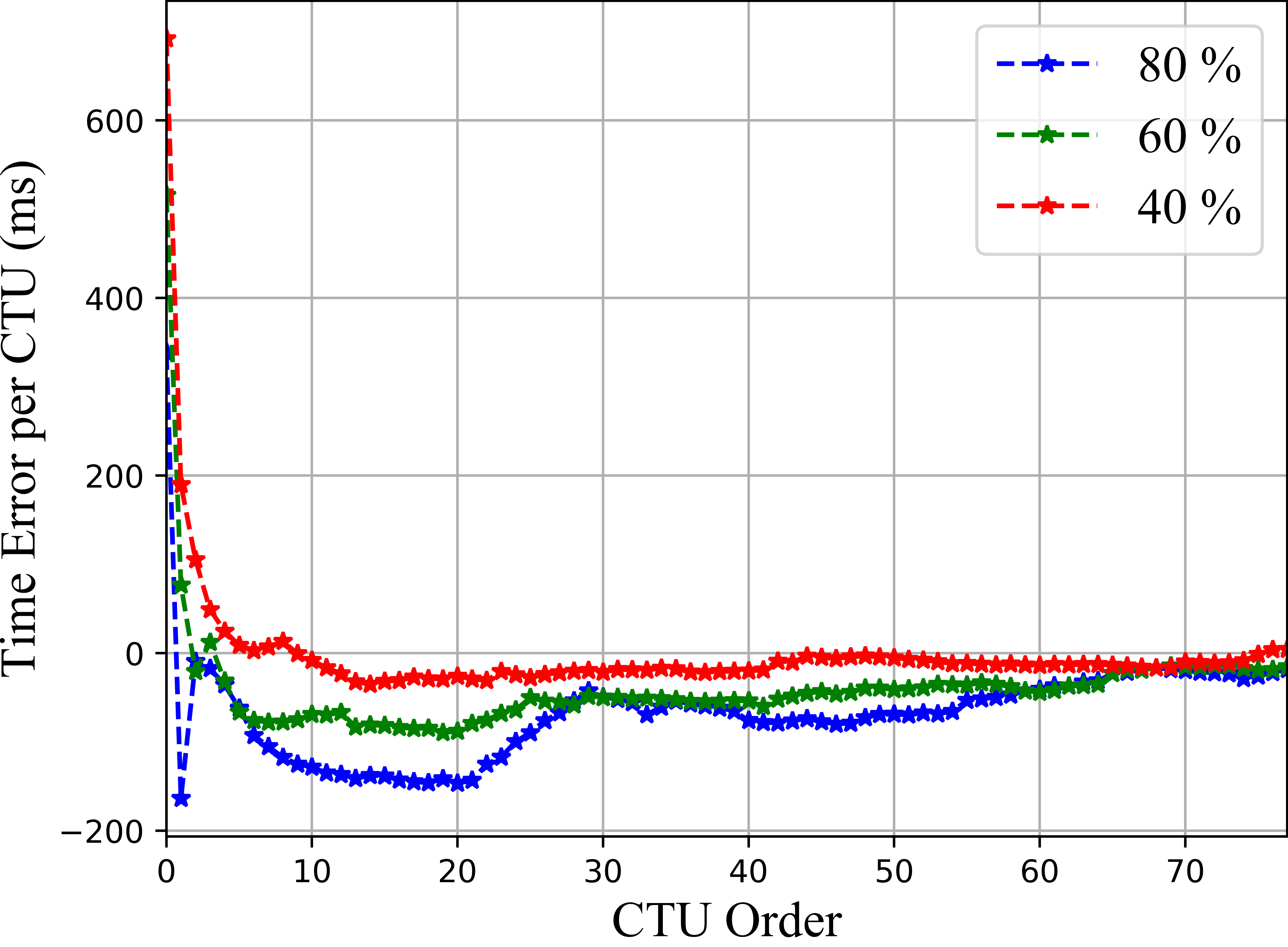}} 
\caption{The time error per CTU as encoding progresses (QP=37)}
  \label{fig:te}
\vspace{-0.4cm}
\end{figure}

Four sequences with different resolutions and three target encoding times (40\%, 60\% and 80\% to original Luma compress time) are presented to show the complexity control effects. Fig. \ref{fig:te} presents the trend of accumulated TE, divided by encoded CTU number.
According to the figures, for sequences of different sizes and target encoding time, the average TE fluctuates at the beginning. But then the framework functions and average TE converges to nearly zero after around a quarter of the picture. This proves the framework has the ability to precisely control encoding time to target.

\subsection{Overall Performance}

\begin{table}[tb]
  \centering
  \caption{Coding Performance under Target Complexity Reduction Ratio}
    \begin{tabular}{cccc}
    \toprule
    \textbf{Target Luma} & \multirow{2}[2]{*}{\textbf{BDBR (\%)}} & \multirow{2}[2]{*}{\textbf{TS}} & \multirow{2}[2]{*}{\textbf{TE}} \\
    \textbf{Compress Time Ratio} &       &       &  \\
    \midrule
    30\%  & 2.71  & 57.78\% & 6.68\% \\
    40\%  & 1.87  & 49.73\% & 3.82\% \\
    50\%  & 1.31  & 41.65\% & 2.57\% \\
    60\%  & 1.00  & 33.73\% & 2.34\% \\
    70\%  & 0.66  & 25.55\% & 2.13\% \\
    80\%  & 0.44  & 17.72\% & 2.27\% \\
    90\%  & 0.23  & 9.96\% & 2.67\% \\
    \midrule
    AVERAGE &       &       & 3.21\% \\
    \bottomrule
    \end{tabular}%
  \label{tab:overall}%
\vspace{-0.4cm}  
\end{table}%

In this part, acceleration properties, i.e., time saving and RD loss, as well as complexity error, are evaluated in terms of TS, BDBR and TE respectively. The same method to last part is used to set target Luma compress time. TABLE \ref{tab:overall} demonstrates the average performance under a Luma compress time corresponding to $30\%\sim90\%$ of the original. 
According to the table, on average 3.21\% TE is achieved, which means the encoding time of Luma compress process is precisely controlled to its target, i.e., from 30\% to 90\%. 
It can be observed that in general, TE is higher when the target time ratio is close to 30\%. This is reasonable because the error ratio is calculated with respect to time budget, and a lower picture budget means a higher TE ratio. In addition, according to TABLE \ref{tab:presets}, when adopting the fastest preset, i.e., preset 5, 28.1\% is an average encoding complexity ratio. There is no guarantee that every sequence can reach lower than 30\%, which could also result in higher complexity error.

From the perspective of total encoding time, from 9.96\% to 57.78\% overall encoding time reduction can be achieved under the BDBR loss from 0.23\% to 2.71\%. As a complexity control method, the RD performance is already comparable to the recently published complexity reduction algorithms.
Notably, the time saving and RD performance is achieved only by constraining the maximum depth of QTMT. If better acceleration strategies can be adopted, possibly better results can be expected.

\section{Conclusions}
This paper proposes a CTU-level complexity control method for the precise picture encoding time control of Luma compress process. Through combining the proposed T-C model and the overall framework, the frame encoding time will precisely converge to any initially set target encoding time achievable. Different from previous work that take the original coding time and target ratio as reference, this work directly controls the actual encoding time, which is more practical in some senarios like achieving a target FPS. To the best of our knowledge, this is the 
first solution targeting complexity control for VVC.
This solution can be regarded as a fundamental method for fine-grained VVC complexity control, and could be further used to control the encoding time of a sequence. T-C model can also be redesigned for the complexity control of VVC Inter coding, which would be our future work.

\section*{Acknowledgment}
This work was supported in part the National Key Research and Development Project of China under Grant 2019YFB1802701, MoE-China Mobile Research Fund Project under Grant MCM20180702, Chinese National Science Funding 62132006, and Shanghai Key Laboratory of Digital Media Processing and Transmissions.
\bibliographystyle{IEEEtran}
\bibliography{ref}

\end{document}